\documentclass[%
 reprint,
 amsmath,amssymb,
 aps,
]{revtex4}

\usepackage{graphicx}
\usepackage{dcolumn}
\usepackage{bm}
\newcommand\dps{\ref@jnl{AAS/DPS Meeting Abstracts}}

\def\ie{{$i.e.$,~}}
\def\eg{{$e.g.$,~}}
\newcommand{\vin}{v_{\rm in}}
\newcommand{\thetain}{\Theta_{\rm in}}

\newcommand{\rin}{r_{\rm in}}
\newcommand{\rci}{r_{\rm ci}}
\newcommand{\rcm}{r_{\rm cm}}
\newcommand{\rco}{r_{\rm co}}
\newcommand{\rc}{r_{\rm c}}
\newcommand{\vc}{v_{\rm c}}
\newcommand{\thetac}{\Theta_{\rm c}}

\usepackage{array,multirow}
\usepackage[]{algorithm2e}
\graphicspath{{./res/}}

\def\lsim{\lower.5ex\hbox{$\; \buildrel < \over \sim \;$}}
\def\gsim{\lower.5ex\hbox{$\; \buildrel > \over \sim \;$}}
\def\be{\begin{equation}}
\def\bea{\begin{eqnarray}}
\def\eea{\end{eqnarray}}
\def\ee{\end{equation}}

\def\mbh{M_*}
\def\rg{r_{\rm g}}

\def\rsh{r_{\rm sh}}

\def\medd{\dot{M}_{\small{\rm Edd}}}

\def\msolar{M_{\odot}}

\newcommand{\qbr}{Q_{\rm br}}

\newcommand{\ndelec}{n_{\rm e^-}}

\newcommand{\ndpro}{n_{\rm p}}

\newcommand{\melec}{m_{\rm {e^-}}}

\newcommand{\kb}{k_{\rm {b}}}

\newcommand{\mpro}{m_{\rm p}}

\newcommand{\mdot}{{\dot{{M}}}}

\newcommand{\rshock}{r_{\rm sh}}
\usepackage{lmodern} 
\usepackage{diagbox}

\newcommand{\trp}{t_{\rm r\phi}}

\newcommand{\as}{a_{\rm s}}
\newcommand{\dqq}{\Delta Q}

\newcommand{\mdotscrinng}{\dot{\cal M}_{r_{\rm in},NG}}
\newcommand{\mdotscring}{\dot{\cal M}_{r_{\rm in},G}}

\newcommand{\tk}{\tilde{K}}
\newcommand{\cmark}{\ding{51}}%
\newcommand{\xmark}{\ding{55}}%
\usepackage{amssymb}
\usepackage{pifont}
\usepackage{tikz}

\begin{document}

\title{Iterative relaxation method to obtain global transonic flows around compact objects}

\author{\firstname{Shilpa}~\surname{Sarkar}}
\email[E-mail: ]{shilpa.sarkar30@gmail.com}
\affiliation{ Harish-Chandra Research Institute, Chattnag Road, Prayagraj, 211019, 	              Uttar Pradesh, India}
\author{\firstname{Igor}~\surname{Kulikov}}
\email[E-mail: ]{kulikov@ssd.sscc.ru}
\affiliation{Institute of Computational Mathematics
and Mathematical Geophysics SB RAS, Lavrentjeva, 6, Novosibirsk Region, 630090, Novosibirsk, Russia}

\received{September 10, 2024; Revised October 20, 2024; Accepted December 2, 2024}

\begin{abstract}
Flows around compact objects are necessarily transonic. Due to their dissipative nature, finding of sonic points is not trivial. \citeauthor{bl03} in 2003 (BL03) proposed a novel methodology to obtain  global transonic solutions, using iterative relaxation technique and exploiting the inner boundary conditions of the central object. 
In the current work, we propose a generic methodology -- IRM-SP and IRM-SHOCK to obtain any class of global accretion and wind solutions, given a set of constants of motion. We have considered viscosity in the system, which transports angular momentum outwards. In addition, it heats the system. Radiative processes like bremsstrahlung which cools the system is also incorporated. An interplay between heating and cooling process, along with gravity and centrifugal forces gives rise to multiple sonic points and hence shocks. The proposed methodology successfully generates any class of accretion as well as wind solutions, allowing us to unify them. 
Additionally, we report here rigorously the mathematical as well as the computational algorithm needed, to find sonic point(s) and thus obtain global transonic flows around compacts objects. 
\end{abstract}

\maketitle

\section{Introduction}

Compact objects are ubiquitous. Due to their large compactness ratios (${\cal C_R}=M_*/R_*$, where $M_*$ and $R_*$ are the mass and radius of the object) they possess strong surface gravitational potentials, and hence, exhibit a phenomenon called \textit{accretion}. This is the process by which the ambient matter is attracted towards the central object and an energy equivalent to the rest mass energy of the infalling matter is emitted in the form of radiation. Thus, if a small mass $\Delta m$ falls onto a star of mass $M_*$, then, $\Delta E=G M_*\Delta m/R_*$, is the amount of energy released due to accretion. The more compact the object (more ${\cal C_R}$), more would be the gravitational potential energy released. Accretion around BHs are known to power the most luminous objects of the Universe which are active galactic nuclei (AGN). Also, they have successfully explained the extreme emissions coming from X-ray binaries (XRBs), gamma-ray bursts (GRBs) and ultra-luminous X-ray sources (ULXs)\cite{king,c96}. Starting in 1939, Bondi-Hoyle-Lyttleton \cite{hl39,bh44,b52} proposed the first-ever models for accretion onto compact stars. \citeauthor{b52}\cite{b52} in 1952, provided us with a more general formulation for obtaining transonic solutions around these objects. Following this, there have been many developments in this field, which targeted to interpret the observations. Out of them, \citeauthor{ss73}\cite{ss73} disk model is one of the most celebrated models used in explaining the high luminosity, soft state of AGNs and XRBs. However, they were found to be thermally and viscously unstable. Therefore, advection-dominated accretion flows (ADAFs) came into picture, where the amount of heat generated in the disk was assumed to be partially advected with the flow, towards the central object. Advection proved to be an important component, providing stability in accretion flows. ADAFs are generally hot accretion flows reaching temperatures of the order of $10^{10-12}$K which could successfully explain the low hard states of the AGNs\cite{sle76,ichi77,abra88,ch96,abra96,chen97,ny94} 
and is presently used to explain the observations of M87 and Sgr A* obtained from the Event Horizon Telescope (EHT) \cite{eht1,eht2,eht3,eht4}. In addition, various other low-luminosity AGNS (LLAGNs) and BH-XRBs have been explained using this model \cite{man97,maha98,nem06,monika09,yn14,bkl01}. However, many of the works present in literature has avoided the transonic feature of accretion flows, \eg in few cases self-similarity was considered \cite{ny95}. These assumptions are arbitrary, since, matter very far away from the central object is subsonic while as it approaches the central object where gravity is strong, matter becomes significantly sonic. Thus, flows around compacts objects are necessarily transonic. Also, transonic solutions have the maximum entropy\cite{b52} and are maximally stable\cite{honma91,chentaam}. In the current work, we focus specifically on finding of global transonic solutions for a large set of parameter space. Few important works done on transonic flows are \cite{lt80,ch96,chen97,f87,kaf94,lee99,bl03,kc14,scp20,skcp23}. These works dealt with different kinds of compact objects as well as gravitational potentials or have assumed different regimes (one-temperature or two-temperature regimes). Even multiple dissipative mechanisms were also considered. Although these works have broadly discussed  the implications of different processes, potentials and regimes, the algorithm to find a transonic solution (TS, hereafter) was never clearly reported. The exact numerical framework was missing. Also, these works focussed more on accretion flows. Few works also had arbitrary conditions imposed at the sonic point (SP, hereafter) in order to obtain TSs. 
Multiple critical point (MCP) regime was not considered in most of the works done in literature. MCP solutions could harbour shocks, and hence, it is important to investigate the whole parameter space. 
Thus, the current work aims to focus more onto obtaining generic TSs, taking recourse to least possible assumptions, broadly discuss the algorithm to obtain global TSs and investigate shocks, for both accretion as well as winds, and thus, present a unified picture of these flows. 

The paper is divided into the following parts: Section 2 presents the model and equations used, Section 3 discusses the methodology and algorithm used to obtain SPs and hence TSs including shocked solutions. In Section 4, we present few results discussing the nature of transonic flows, and at the end, in Section 5, we conclude the work by discussing the possible implications and caveats present in the current problem.

\section{Modelling and Equations used}

We consider steady, axisymmetric, viscous, dissipative flow, assuming \citeauthor{pw80} \cite{pw80} (hereafter, PW) potential, which mimics the effects of general relativity. Hydrostatic equilibrium is considered in the perpendicular direction, to the plane of the flow. Unit system considered is $2G=\mbh=c=1$ (where $G$ is the gravitational constant and $c$ is the speed of light in vacuum), such that velocity, length and time are in units of $c$, $\rg=2G\mbh/c^2$ and  $2G\mbh/c^3$ respectively. Since we are dealing with trans-relativistic flows, the adiabatic index ($\Gamma$) would vary between 5/3 (non-relativistic regime) to 4/3 (relativistic regime). Thus, we use an equation of state (EoS) with variable adiabatic index and multi-species component information which was given by \citeauthor{cr09}\cite{cr09}. The EoS has the following form:
\begin{equation}
e= \dfrac{\rho f}{\tk},
\label{eq:eos}
\end{equation}
\begin{equation}
{\rm where,~}f = (2-\xi) \left[ 1+\Theta \left( \frac{9 \Theta +3}{3 \Theta +2}\right)\right]+\xi \left[ \frac{1}{\chi}+\Theta \left( \frac{9 \Theta +3/\chi}{3 \Theta +2/\chi}\right)\right] ~~ {\rm and,~}\tk=2-\xi(1-1/\chi).\nonumber
\end{equation}
Here, $\rho$ is the total mass density of the species, $\xi=\ndpro/\ndelec$ is the composition parameter, $\chi=\melec/\mpro$ and $\Theta=\kb T/(\melec c^2)$ is the dimensionless temperature defined w.r.t the rest mass energy of the electron, $\kb$ being the  Boltzmann constant. The adiabatic index can then be defined as, $\Gamma=1+1/N$ where, $N=(1/2)~df/d\Theta$ is the polytropic index. We have assumed charge neutrality in the system. 
The system is of one-temperature ($T=\Theta \melec c^2/\kb$) \ie all the species have settled down into a single temperature distribution. This assumption relieves us from the complexities present in two-temperature theory \citep{sc19ijmpd, scp20, sc19jp, sc22jaa, skcp23}. 
We enumerate below the conservation equations, simplified using the EoS (Eq.~\ref{eq:eos}) and assuming the PW potential. 

\begin{enumerate}
\item The continuity equation can be integrated to obtain a constant of motion for mass flux known as the mass accretion rate ($\mdot$) which is given as:
\begin{equation}
\mdot=2\pi r\rho v 2H =2\pi rv\Sigma.
\label{eq:mdotdisc}
\end{equation}
Here, $v$ is the radial velocity, $r$ is the radial coordinate, $H$ is the half-height of the flow\cite{kc14}  and $\Sigma=2\rho H$ is the height-integrated density. $\mdot$ is a constant of motion throughout the flow in the absence of pair production and annihilation, which is the case in the current work\cite{sc20jp}. 
\item The azimuthal component of the the momentum balance equation is:
\begin{equation}
\frac{d\lambda}{dr}+\frac{1}{\Sigma v r}\frac{d(r^2 \trp)}{dr}=0,
\label{eq:lambdaeq1}
\end{equation}
where, $\trp$ is the viscous stress. Integrating the above equation using Eq.~\ref{eq:mdotdisc}, we get:
\begin{equation}
\dot{M}(\lambda-\lambda_0)=-2\pi r^2 \trp,
\label{eq:lambdaeq2}
\end{equation}
where, $\lambda_0$ is the specific angular momentum at the inner boundary (which is horizon in case of BHs)\citep{bl03,bs05}. 
We follow the $\trp \propto \alpha p$ viscosity prescription in the current work, where $\alpha$ is the Shakura-Sunyaev viscosity parameter \citep{ss73}. This is famously used by many authors, because it reduces Eqs.~\ref{eq:lambdaeq1}--\ref{eq:lambdaeq2} to an analytical form \citep{sc22jaa, sado09,mat84}, expression of which is given below:
\begin{equation}
\lambda=\lambda_0+\frac{2\pi r^2 \alpha W}{\dot{M}}.
\label{eq:ang}
\end{equation}
where, $W$ is the height-integrated gas pressure $W=2Hp$.

\item The first law of thermodynamics is given by:
\begin{equation}
\frac{d\Theta}{dr}=-\frac{2\Theta}{2N+1}\left[ \frac{1}{v}\frac{dv}{dr}+\frac{5r-3}{2r(r-1)}\right]-\frac{\dqq \tk}{\rho v (2N+1)},
\label{eq:flt}
\end{equation}
where, $\Delta Q=Q^+-Q^-$ is the dissipation present in the system where, $Q^+$ is the heating \cite{mat84,naka96,sudeb22,sc22jaa} while $Q^-$ denotes the cooling.
To keep the model simple, we have assumed only bremsstrahlung as a cooling term \citep{scp20, skcp23}. 
\item The radial component of momentum balance equation can be written as:
\begin{equation}
v\frac{dv}{dr}+\frac{1}{\rho}\frac{dp}{dr}-\frac{\lambda^2}{r^3}+\frac{1}{2(r-1)^2}=0.
\label{eq:dvdrdisc1}
\end{equation}
The differential equation for velocity is obtained by simplifying the above equation using Eqs.~\ref{eq:eos}, \ref{eq:mdotdisc} and \ref{eq:flt} and is given by:
\begin{equation}
\dfrac{dv}{dr}=\dfrac{\as^2\left[\dfrac{2N}{2N+1} \dfrac{5r-3}{2r(r-1)}\right]+\dfrac{\lambda^2}{r^3}-\dfrac{1}{2(r-1)^2}+\dfrac{\dqq}{\rho v (2N+1)}}{v \left[1 -\dfrac{\as^2}{v^2}\left( \dfrac{2N}{2N+1}\right)\right]}=\cfrac{\cal N}{\cal D},
\label{eq:dvdrdisc2}
\end{equation}
where, $\as^2=2\Gamma \Theta/\tk$ is the sound speed. At some point of the flow, ${\cal D}=0$. For the flow to be continuous, ${\cal N}$ also has to go to 0. This point is called the critical point (CP) of the flow while SP is where $v=\as$. Due to the assumption of hydrostatic equilibrium CP$\ne$SP\cite{susovan22}.

\item Bernoulli parameter: On integrating Eq.~\ref{eq:dvdrdisc1} we get a constant of motion called the  generalised Bernoulli constant for viscous radiative flows \citep{bl03,kc14}:
\begin{equation}
\dfrac{1}{2}v^2+h-\dfrac{\lambda^2}{2r^2}+\dfrac{\lambda \lambda_0}{r^2}-\dfrac{1}{2(r-1)}-{\cal Q}={E},
\label{eq:bernoullidisc}
\end{equation}
where, $h$ is the enthalpy and ${\cal Q}=\int [\qbr /(\rho v) ]dr$. 
In the absence of any dissipation, we retrieve the canonical form of Bernoulli constant which is given as:
\begin{equation}
\dfrac{1}{2}v^2+h+\dfrac{\lambda_0^2}{2r^2}-\dfrac{1}{2(r-1)}={\cal E}.
\label{eq:bernoullican}
\end{equation}
\item Entropy accretion rate: Integrating the first law of thermodynamics (Eq.~\ref{eq:flt}) in the adiabatic limit gives an expression for entropy accretion rate:
\begin{equation}
{\cal \mdot}=vHr\Theta^{3/2}(3\Theta+2)^{3(2-\xi)/4}(3\Theta+2/\eta)^{3\xi/4}\exp[(f-\tk)/(2\Theta)].
\label{eq:eac}
\end{equation}
The above expression is not a constant of motion in the current work, since dissipation is involved in the system. We would use this formula only in regions where it is locally adiabatic. 
\end{enumerate}

\subsection{Sonic points and their properties}
SPs are formed when the velocity of the infalling matter crosses the speed of sound in that medium.  We define a quantity called Mach number ($M$) which is the ratio between these values or $M=v/\as$. In case of spherical (or Bondi\cite{b52}) flows, a single sonic point is formed due to gravity. 
 When angular momentum is present, due to rotation of the matter, multiple critical points may form inside the flow. 
 The SPs are named according to their distance from the central object: inner ($\rci$), middle ($\rcm$) and outer ($\rco$). While $\rci$ and $\rco$ are saddle or X-type SPs, \ie matter actually passes through them; $\rcm$ is an O-type SP which is center/spiral-type in the absence/presence of dissipation respectively, and hence, matter do not physically pass through it. Thus, to obtain a TS, we need to locate $\rci$ or $\rco$ or both, depending on the assumed set of flow variables. It is to note that in the current work, we have used the terminology CP and SP analogously. The finding of SP signifies finding the CP only.
\subsection{Shocks}
\label{sec:shocks}
In the presence of MCP where 2 X-type SPs are present, shocks could be formed. In the presence of shocks, subsonic matter starting from infinity (for accretion flows) would first pass through $\rco$ and become significantly sonic 
$\rightarrow$ encounter a shock transition $\rightarrow$ become subsonic $\rightarrow$ again become significantly sonic, after passing through $\rci$. For the case of winds, the subsonic flow will start from the surface and flow a similar path. In the absence of shocks, but in MCP regime, the flow will remain significantly sonic throughout the upstream region. 
Location of shock could be determined from Rankine-Hugoniot (RH) conditions, \ie mass flux, energy flux amd radial as well as azimuthal component of momentum flux conservation\cite{ll59}. The conditions are respectively: (i) $\mdot_+=\mdot_-$, (ii) $\dot{E}_+=\dot{E}_-$, (iii)  $\dot{J}_+=\dot{J}_-$ and (iv) $W_++\Sigma_+v_+^2=W_-+\Sigma_-v_-^2$. Here, $-$ and $+$ denote the quantities before and after the shock transition and $\dot{J}=\mdot\lambda+r^2\trp=\mdot \lambda_0$. 


\section{Algorithm and Computational Scheme}
The methodology used to solve the above set of equations or the equations of motion (EoM) is inspired from  \citeauthor{bl03} (2003) (BL03 hereafter) , who used the inner boundary conditions of the BH event horizon to obtain accretion solutions. An iterative relaxation method was employed, alongwith the stress-free condition, which is valid only at the horizon and not at the SP\cite{gn92}. 
This work is hence, one of the pioneering works related to obtaining of global transonic accretion solutions around compact objects. Thereafter, many works have employed this methodology for modelling flows around BHs\cite{kc14,scp20,mitra23} as well as NSs\cite{skcp23}. However, these works have majorly discussed accretion solutions and also avoided MCP regime harbouring shocks (\eg BL03). 
In this section, we investigate the whole parameter space. The exact mathematical treatment and computational scheme employed to get the location of SP and hence the TS and shock, would be discussed. 
We divide this section into two parts, (1) discuss the methodology adopted for obtaining TSs that are connected with the inner boundary, or in other words, by using inner boundary conditions we find the SPs and hence obtain TSs (method name: IRM-SP), and (2) propose an extension of the IRM-SP methodology to obtain TSs which are not  connected with the compact object. These solutions would need shock transitions to get connected to the inner boundary (method name: IRM-SHOCK).


\subsection{Transonic solutions \textbf{connected} with the central object (Method name: IRM-SP)}

The class of solutions which could be retracted using this method are:
\begin{itemize}
\item \textit{All accretion solutions}: Accretion solutions harbouring single or MCP are all connected to the horizon. Accretion solutions could be global: connecting horizon to infinity through a single SP 
or could be non-global: connecting horizon upto a certain radius after passing through $\rci$. 
 In the MCP regime (both $\rci$ and $\rco$ are present), the global solutions passing through $\rco$, could connect to the non-global ones (passing through $\rci$)  via a shock transition, only when the RH conditions are satisfied. 
\item \textit{Wind solutions passing through single SPs}: This methodology allows us to obtain global wind solutions, that directly connect the central object to infinity, passing through a single SP. 
\end{itemize}

All the above set of solutions could be obtained using the methodology described below. Hereafter, we name this methodology as IRM-SP, which stands for iterative relaxation method used for obtaining SPs. We present this graphically in Fig.~\ref{fig:isp}.
\begin{enumerate}
\item \textbf{FLOW PARAMETERS}: Supply $E$, $\lambda_0$, $\alpha$, $\mdot$, $\mbh$ and $\xi$.
\item \textbf{INNER BOUNDARY VALUES}: Select a boundary and branch: We choose $\rin \rightarrow \rg =1.001$ to be the  boundary and assume the accretion branch. The reason for these assumptions are because, very close to the central object, where gravity is strong, accretion is faster or infall timescales are much shorter than any other timescales. Hence, $E\simeq {\cal E}$ (see, Eq.~\ref{eq:bernoullican}), and we have an analytical expression for energy. Simplifying this equation, we get: $\vin=\sqrt{2[{E}-h(\thetain)-\lambda_0^2/(2\rin^2)+1/{2(\rin-1)}]}$. Here, $h$ is a function of only $\thetain$. In the current expression to obtain $\vin$, we have assumed $\lambda_{\rm in}\approx\lambda_0$, since their difference is generally less than $10^{-10}$\cite{bl03}. 
We supply a guess value of $\thetain$ and obtain $\vin$, from the aforementioned expression.
\item \textbf{IRM-SP TECHNIQUE}: Using $\rin$, $\vin$ and $\thetain$ at $\rin$ as boundary conditions, we employ explicit adaptive Runge-Kutta (RK) method to solve the set of coupled differential equations (the EoM) \ie $dv/dr$ (Eq.\ref{eq:dvdrdisc2}), $d\theta/dr$ (Eq.\ref{eq:flt}) and the analytical expression for $\lambda$ (Eq.~\ref{eq:ang}). We use the  Butcher tableau of Cash-Karp RK method which is $5^{\rm th}$ order accurate\cite{ck}.
The solution obtained, may not be transonic or do not satisfy CP conditions (${\cal N}={\cal D}=0$) at any $r$. Hence, an iterative relaxation method\citep{press} needs to be utilized. The value of $\thetain$ is thus iterated, unless at any $r=\rc$, the CP conditions are satisfied. In Fig.~\ref{fig:isp}~(a1) and (b1) we plot the different solutions represented using $M$ and $v$ respectively, for different values of $\thetain$ (and correspondingly $\vin$, step 2). The flow parameters assumed are $E=1.001,~\lambda=1.5,~\alpha=0.01,~\mdot=0.1\medd,~\mbh=10\msolar, ~\xi=1.0$. We discuss below the computational scheme for the relaxation method. 
\begin{figure}[!htb]
\centering
\includegraphics[scale=0.7,  trim={3.2cm 6.8cm 4cm 8.cm},clip]{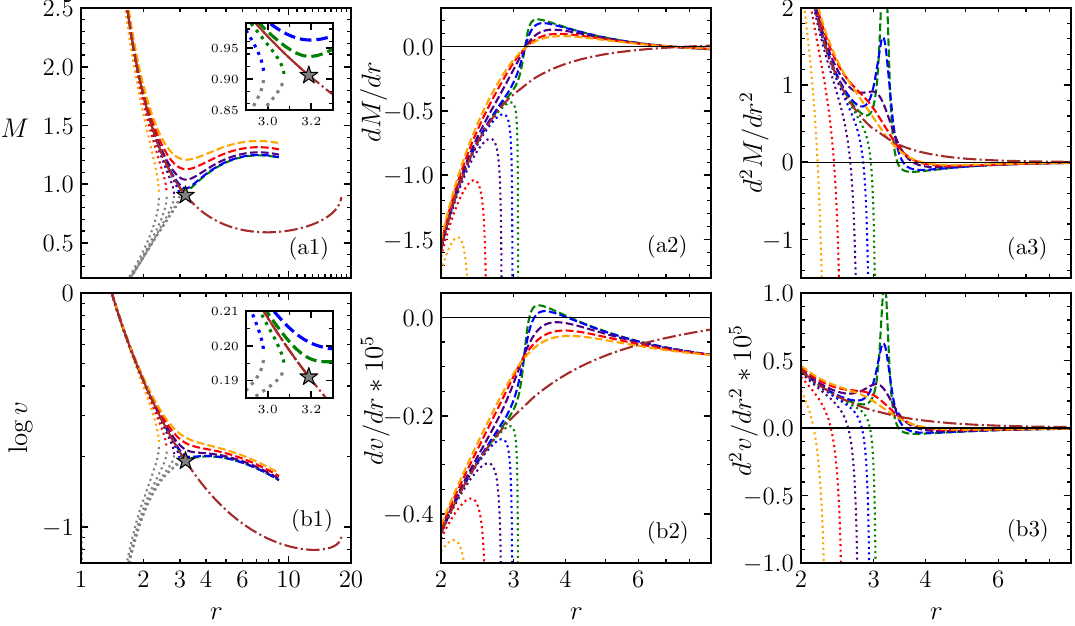}
\caption{IRM-SP used to obtain inner SP. Panels (a1) and (b1) plot the $M$ and $v$ respectively w.r.t $r$ for different values of $\thetain$. Dashed curves represent $\thetain$s leading to significantly sonic branch (SB) solutions, while dotted represents multi-valued branch (MVB) solutions. The transonic solution (TS) is represented using dashed-dot brown curve with the sonic point marked using grey star. Panels (a2-a3) and (b1-b3) plots the different diagnostics which could be used to identify the different curve topologies (going up, SB or going down, MVB), also see table \ref{tab:1}. The flow parameters  are: $E=1.001,~\lambda=1.5,~\alpha=0.01,~\mdot=0.1\medd,~\mbh=10\msolar, ~\xi=1.0$. }
\label{fig:isp}
\end{figure}
\begin{enumerate}
\item Given a $\thetain$ (and correspondingly obtain $\vin$), we get a solution. Let $\thetain=130.459$ and we get the yellow dashed curve (see, Fig.~\ref{fig:isp} (a1)). This branch is significantly sonic throughout the flow (named as SB hereafter). On the other hand, if $\thetain=141.161$, we get the yellow dotted curve which harbours two values of $M$ (or $v$), or in other words, is multi-valued, at any given $r$ (named as MVB hereafter)\footnotetext{While finding the MVB only the coloured dotted curves could be obtained, after which the integration is not possible. This is due to  computational limitation. To obtain the full branch, $v$ and $\Theta$ are reduced by a factor of $<10^{-6}$, such that it can jump to the subsonic branch. Thereafter, the integration is done in the negative $r$ direction. This subsonic part is represented using grey dotted curves. The combination of the two curves (coloured + grey) represents the full solution which is here MV. Even without generating the subsonic branch, the diagnostics could be used.}. Both these branches are unphysical. However, the TS lies in between these two $\Theta$ limits.
\item To identify whether the solution is going up (SB) or down (MVB) we have 4 different diagnostics: $dM/dr$, $d^2M/dr^2$, $dv/dr$ and $d^2v/dr^2$, see, table \ref{tab:1}. The tick mark in the table denotes whether the diagnostic could be used to identify the type of branch and cross represents just the opposite. Degenerate means that, no conclusive result could be drawn, since the diagnostic prompts positive irrespective of the nature of branch.
\begin{table}[!htb]
    \begin{tabular}{|c|c|c|c|c|}
        \hline
\diagbox[width=\dimexpr 1.7\textwidth/10+0\tabcolsep\relax, height=1.3cm]{ Branch }{Diagnostics}     
       
         & $dM/dr$ &$d^2M/dr^2$& $dv/dr$ & $d^2v/dr^2$ \\
        \hline\hline
        Up &  \cmark & \xmark & Degenerate & \xmark \\
        \hline
        Down &  \xmark & \cmark & \xmark & \cmark \\
        \hline
    \end{tabular}
    \caption{Usable diagnostics for identifying the significantly sonic (up) and multi-valued (down) branch}
        \label{tab:1}
\end{table}
\item To check whether the solution is going up only $dM/dr$ could be used \ie if there is an extrema in $M$ w.r.t $r$, then definitely the solution is SB. All the dashed curves pass through 0 in panel (a2). $dv/dr$ could not be used, since few SB curves harbour an extrema in $v$ while few do not, and hence a degeneracy prevails (see panel b2). $d^2M/dr^2$ and $d^2v/dr^2$ do not provide us with any usable information regarding the nature of the solution.
\item For solutions corresponding to MVB (going down, dotted curves), there is a maxima in $dM/dr$ (panel a2) as well as $dv/dr$ (panel b2). This maxima can be located using their double derivatives. Thus, MVB can be identified if either $d^2M/dr^2$ (panel a3) or $d^2v/dr^2$ (panel a4) changes sign. 

\item We iterate $\thetain$, identify the direction of branches and finally reduce $\thetain$ limits to the lowest possible values. From Fig.~\ref{fig:isp}~(a1, b1), we see that, from yellow dashed (SB) and dotted (MVB) curves, $\thetain$ limits are reduced to  red curves, then to blue, indigo and finally green curves. Thereafter, we achieve upto computer precision, the value of $\thetain=134.333$ for which CP conditions are satisfied at $r=\rc=3.188$ (grey star) and the hence the TS is obtained (dashed-dot brown curve).

\end{enumerate}
\item \textbf{OBTAINING TS}: After, $\rc$, $\vc$ and $\thetac$ are obtained, we integrate the EoM, inwards (negative $r$) and outwards (positive $r$). However, $\rc$ being a critical point,  the expression of $dv/dr|_{\rc}$ needs to be derived using L'Hopital's rule, which is given by, $(d{\cal N}/dr)/(d{\cal D}/dr)$. Simplifying this equation gives two values of $dv/dr$ at $r=\rc$. Using the negative $dv/dr$ would give the accretion solution, while the positive would provide us with the wind solution. Fig.\ref{fig:osp} (a--c), dashed-dot brown curve plots the solution with negative $dv/dr$ at $\rc$, while dotted black curve plots TS with positive $dv/dr$. For the current set of flow parameters chosen, we see that the TS passes through $\rci$ and is non-global, \ie does not connect the outer boundary with the inner boundary.

\item \textbf{MCP}: To check whether there are any other SPs present, we reduce the $\thetain$ (used to obtain SP in step 3) by a large factor and follow the same IRM-SP technique (step 3), keeping the flow parameters (step 1) unchanged. However, we need to apply the diagnostics only after the $\rci$ location. Fig.~\ref{fig:osp} (a1) plots the IRM-SP used to obtain $\rco$, which is exactly same as finding $\rci$ in Fig.~\ref{fig:isp} (a1). The SP is marked using grey dot. Following step 4, we obtain the TS. For the present case, the TS passing through $\rco$ is global (G, hereafter), in contrast with the non-global (NG, hereafter) TS passing through $\rci$. Accretion branch is plotted using solid brown curve while the wind branch using dashed magenta curve, see Fig.~\ref{fig:osp} (b).
It is to note that, the current set of flow parameters harbour MCPs. However, if the parameters are changed, then MCPs may vanish. 
\begin{figure}[h]
\centering
\includegraphics[scale=0.7,  trim={3.4cm 11.5cm 6.8cm 7.6cm},clip]{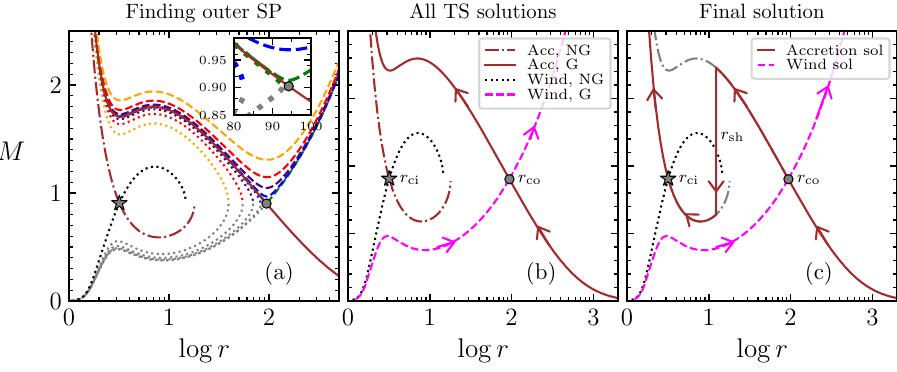}
\caption{Panel (a) plots IRM-SP technique used to obtain outer SP, (b) plots all the TSs present and (c) global accretion (solid, brown) and wind solution (dashed, magenta). The current set of flow parameters harbour shock. $\rci$ and $\rco$ are the location of inner and outer SPs, while $\rshock$ is the location of the shock. The flow parameters are same as that  in Fig.~\ref{fig:isp}. }
\label{fig:osp}
\end{figure}
\vspace{-0.5cm}
\item \textbf{SHOCK CONDITIONS} (\textit{optional; used only in MCP regime}): For the case where MCPs are present, there is a possibility of shock transition from the  global branch (Acc, G; solid brown) to the non-global branch (Acc, NG; dashed-dot brown in Fig.~\ref{fig:osp} (b)). The location, where the radial momentum flux matches (see, section \ref{sec:shocks}), is the location of the shock. It is to note that, the other shock conditions are already satisfied in the two branches and have been already used, to obtain these solutions, \ie $\mdot$ and $\dot{E}=E/\mdot$ are constants of motion throughout the flow, while $\lambda_0$ is fixed at the inner boundary, which is equivalent to the conservation of momentum-flux in the $\phi$-direction. 
We see in the upcoming section that this simple method becomes complex in case of winds and determining of shock locations is not trivial.
For the present set of flow parameters, the momentum flux is equal at two locations, suggesting two permissible shock jumps. However, the inner shock (shock near the central object) is unstable to perturbations\cite{c96}. The only stable shock present is at $\rshock=12.281$. Also, the entropy accretion rate at $\rin$ (Eq.~\ref{eq:eac}), $\mdotscrinng>\mdotscring$. Thus, the shock jump is permissible, since nature prefers a state with maximum entropy.

\item \textbf{FINAL SOLUTION:} In the presence of single SPs, the final TS is the solution connecting infinity to the central object (through $\rco$ or $\rci$). But, in the case of MCPs,  shock could be present. If RH conditions are satisfied at any given $\rshock$, then the global solution would pass through both the SP via a shock transition in between ($\rci<\rsh<\rco$). 
For the present set of flow parameters used, the global accretion solution encounters a shock and is plotted using solid brown curve in Fig.~\ref{fig:osp} (c). The global wind solution is plotted using dashed magenta line.
\end{enumerate}

\subsection{Transonic solutions \textbf{disconnected} with central object (Method name: IRM-SHOCK)}
\begin{figure}[htb!]
\centering
\includegraphics[scale=0.7,  trim={3.4cm 11.5cm 6.8cm 7.6cm},clip]{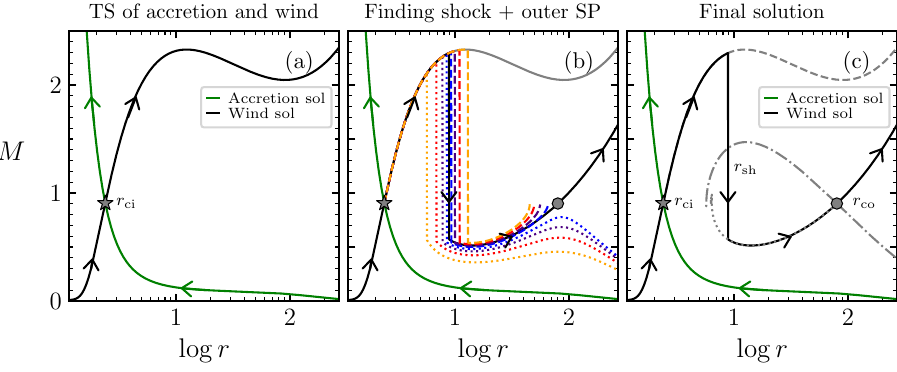}
\caption{(a) IRM-SP used to obtain TS passing through $\rci$, (b) IRM-SHOCK technique utilised to obtain $\rco$ and therefore $\rshock$, (c) global accretion (green curve) and wind solution (black curve)}
\label{fig:wind}
\end{figure}

Unlike non-global TS passing through $\rci$ (see, Fig.~\ref{fig:osp} (b)), NG-TS passing through $\rco$ are not connected to the inner boundary (central object) (see Fig.~\ref{fig:wind} (c)). Hence, IRM-SP technique cannot be employed.  It is to remember that the importance of NG-TS passing through $\rco$ is for wind branch only. 
Global accretion solution passing through $\rci$ (green curve in Fig.~\ref{fig:wind} (c)) possess higher entropy than the NG-TS branch passing through $\rco$ (dashed-dot grey curve in Fig.~\ref{fig:wind} (c)). Thus, shock transition is not possible. However, in case of winds, the shock jump is permissible since the entropy of the global transonic wind solution (passing through $\rci$, black curve in Fig.~\ref{fig:wind} (a)) < non-global transonic wind branch passing through $\rco$ (dotted grey curve in Fig.~\ref{fig:wind} (c)). Utilising this fact, we can locate $\rco$. If the wind branch do not possess such shock jump, then locating $\rco$ is not needed since it never contributes to the global solution.

We use a different set of flow parameters for explaining this methodology named as IRM-SHOCK. We use $\lambda=1.7$, while keeping all other parameters of the previous section unchanged. The different steps adopted in this methodology are enumerated below and is graphically represented in Fig.~\ref{fig:wind}.

\vspace{-0.5cm}
\begin{enumerate}
\item \textbf{OBTAINING GLOBAL TS PASSING THROUGH $\rci$}: First step is to locate $\rci$ and obtain the TS (accretion and wind). For this purpose, we use the ITR-SP methodology discussed in the previous section, and follow steps 1--4. For the new set of flow variables, the global accretion (solid green) and wind solution (solid black) is presented in Fig.~\ref{fig:wind} (a). The SP is marked using grey star.
\item \textbf{OBTAINING $\rshock$ FOR WIND}:  On simplifying the 4 RH conditions (see, Section \ref{sec:shocks}), expressions of $v_+, \Theta_+$, $\lambda_+$ and $\rho_+$ can be obtained as functions of the pre-shock values\cite{kc14}. For every $r_i>\rci$, we force a shock jump, \ie by using $r_-=r_i, ~v_-=v_i,~\lambda_-=\lambda_i$ and $\rho_-=\rho_i$, we get the post-shock values. Then, we integrate the EoM, using the post-shock values as boundary, to obtain a solution. 
However, this solution need not be transonic. Hence, we utilise the same IRM technique to obtain the $\rshock$; iterating on $r_i$ instead of $\thetain$.
In Fig.~\ref{fig:wind} (b), we can see dotted/dashed curves undergoing shock transitions at different $r_i$s. Out of these transitions, only 1 solution passes through the SP (grey dot)  represented by solid black curve. The $r$, for which the solution passes through $\rco$ is the location of $\rshock$ for the assumed set of flow parameters. Thus, we obtain $\rco$ as well as $\rsh$ at the same time.
\item \textbf{GLOBAL SOLUTION}: In the absence of wind shock, IRM-SP method is sufficient to identify all the SPs and the TS. In case wind shock is present, the global solution would pass through $\rci$ and then through $\rco$ via a shock transition at $\rshock$. Global accretion solution would pass through a single $\rci$ (see, Fig.~\ref{fig:wind} (c)). 
\end{enumerate}

\section{Analysis of the solutions obtained}
We discuss and analyse here the typical solutions obtained using the two techniques proposed before. We have fixed $\alpha=0.01$, $\mdot=0.1\medd$, $\mbh=10\msolar$ and $\xi=1.0$.
\begin{figure}[htb!]
\centering
\includegraphics[scale=0.6,  trim={3.2cm 2.9cm 1cm 1.9cm},clip]{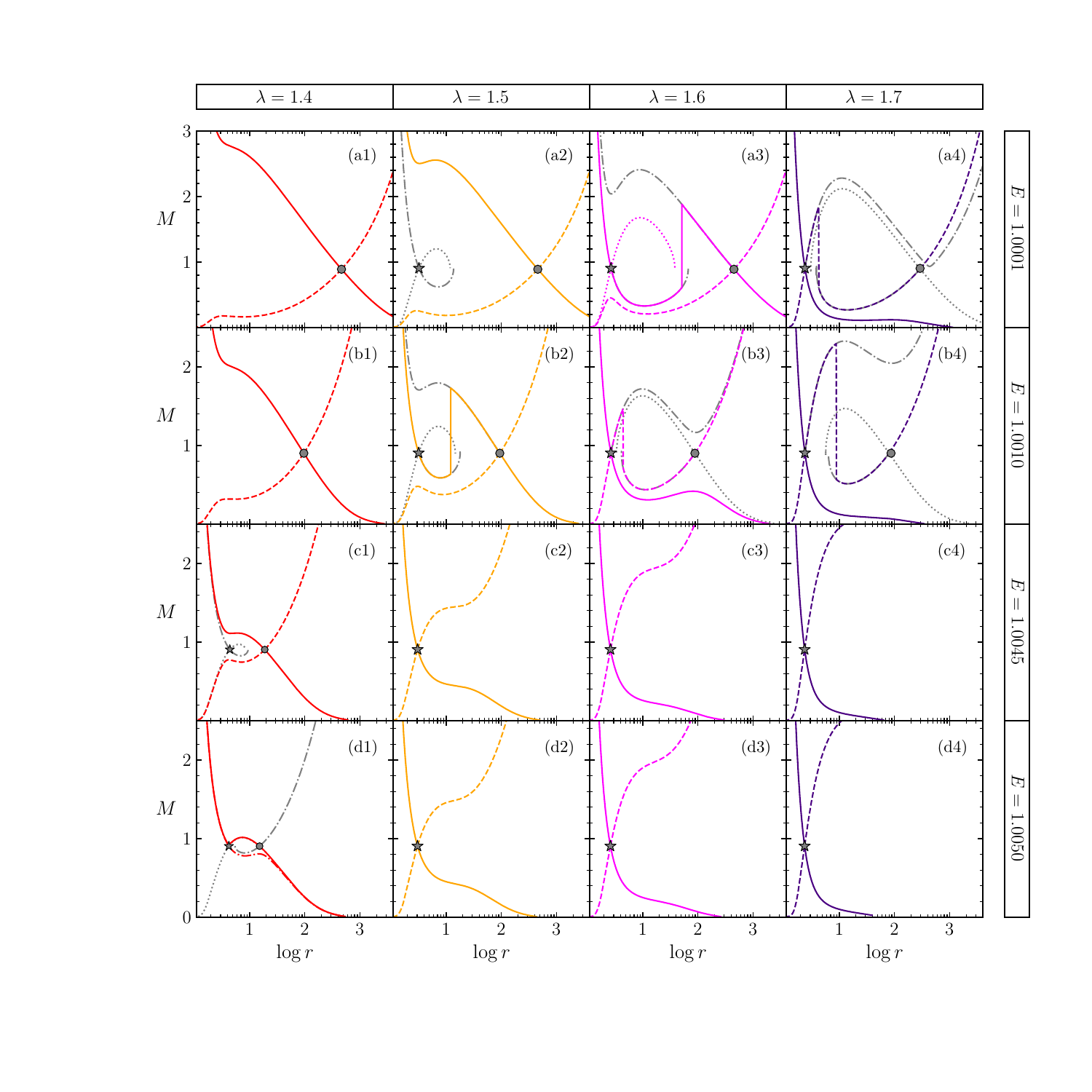}
\caption{Variation of solutions with change in $E$ and $\lambda$ (values written inset). $\rci$ and $\rco$ are represented using grey star and grey circle respectively. Solid curves are global accretion solutions while dashed curves are global wind solutions. Grey curves represent non-global or truncated solutions because of sudden shock jump. Panels (a3) and (b2) harbour accretion shocks while (a4), (b3) and (b4) harbour wind shocks.}
\label{fig:elam}
\end{figure}
\subsection{Energy-angular momentum ($E-\lambda$) parameter space}
In Fig.~\ref{fig:elam}, we plot the different topologies of solution as a  function of $E$ and $\lambda$. 
From left to right, we have increased $\lambda$ and from top to bottom $E$ is increased (values given in figure). Solid curves represent accretion solutions while dashed curves are wind solutions. The grey curves represent NG solutions or solutions not followed because of a sudden shock jump. Grey star and grey circle denotes the $\rci$ and $\rco$ respectively. IRM-SP method is employed to obtain all the solutions connected with the central object. So, majority of the solutions have been generated using this method. IRM-SHOCK is used to obtain $\rco$ where the TS is non-global, \ie in panels (a4), (b3) and (b4) (grey curves and coloured dashed curves passing through $\rco$). Because of the presence of wind shock, the $\rco$ could be determined in these cases. 

For very low or very high $E$ and $\lambda$, the solution harbours a single SP. This is due to the extreme gravity dominating centrifugal force (formation of $\rco$) and vice-versa. For intermediate values of $E-\lambda$ where the gravity and centrifugal force becomes comparable, MCPs are produced. In few cases this gives rise to accretion and wind shocks as well, see, panels (a3) and (b2) for accretion shocks and (a4), (b3) and (b4) for wind shocks.

\subsection{Accretion and wind flow properties}
\begin{figure}[htb!]
\centering
\includegraphics[scale=0.55,  trim={0.2cm 1.2cm 0.1cm 13.cm},clip]{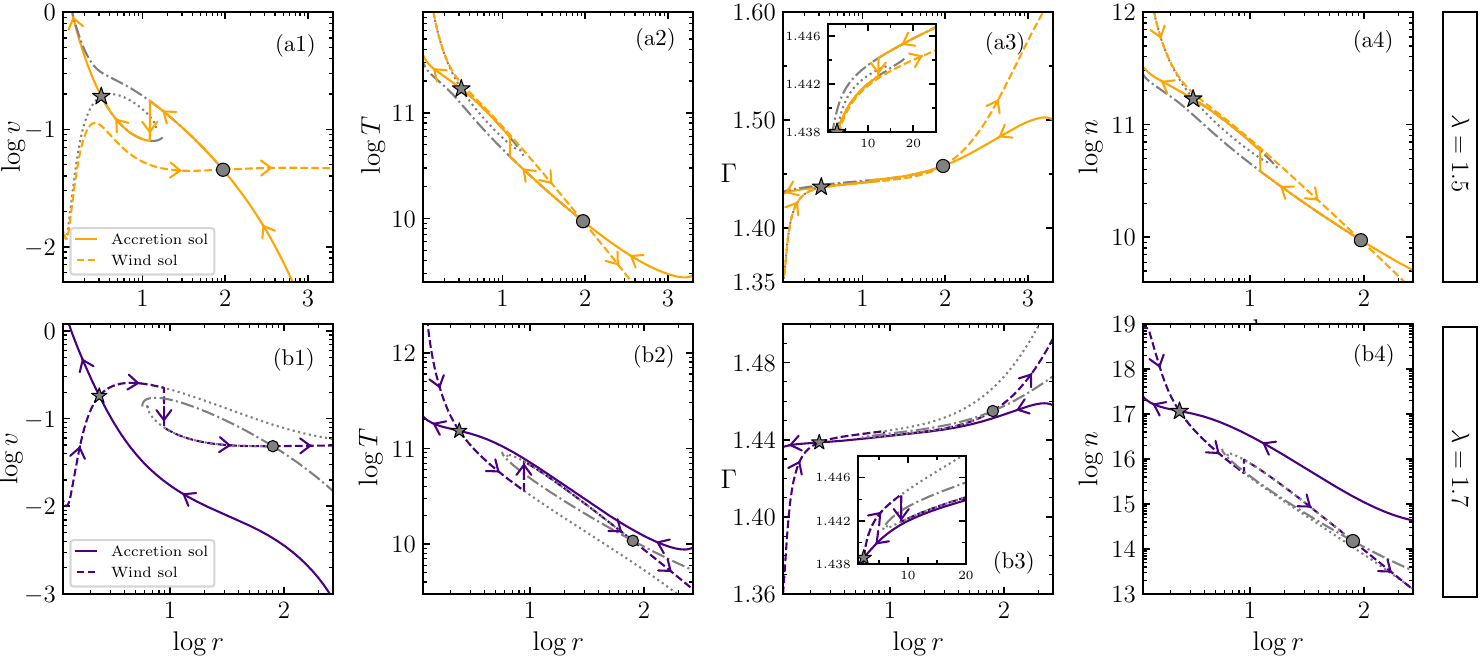}
\caption{Flow variables for solutions represented in panel (b2) and (b4) of Fig.~\ref{fig:elam}.  Coloured curves are the global transonic solutions. Arrows represent the direction of the flow.}
\label{fig:sol}
\end{figure}

We select solutions plotted in panels of (b2) and (b4) of Fig.~\ref{fig:elam} and analyse them. 
In Fig.~\ref{fig:sol}, we plot  their respective $v$, $T$, $\Gamma$ and $n$ ($=\rho/(\melec+\mpro)$). It is seen that, both these flows harbour MCPs as well as shocks. While the former harbours an accretion shock the latter has a wind shock. In both the cases, solid lines represent accretion while dashed represents wind solution. For accretion solution, the direction of flow is from infinity to the surface, while it is just the reverse in case of winds (arrows in the figure). For accretion branch of $\lambda=1.5$ or 1.7, radial velocity ($v$) is very low at infinity, while, it is significantly sonic near the central object (solid curves in panels a1, b1). Temperatures increase by  2-2.5 orders of magnitude due to accretion (panels a2, b2, solid curves) and a corresponding change in $\Gamma$ is observed. Since, we have used CR EoS, the $\Gamma$ is not fixed but varies with $T$ (panels a3, b3, solid curves). For the present set of flow parameters, both the accretion solutions are mildly relativistic throughout the flow. However, for $\lambda=1.5$, where the $T$ change is higher, the  $\Gamma$  change is also higher, as compared to the solution with $\lambda=1.7$. Wind solutions show the reverse nature, it starts with subsonic velocities and reach significantly sonic values as it goes farther and farther out (dashed curves in a1, b1). However, after a certain radius, the change in $v$ is not sharp as was in the case of accretion. The speed hovers around a particular value. The change in $T$ is however very large and is $>3-4$ orders of magnitude (dashed curves in panels a2, b2). The sharp decrease in temperature at large $r$, is due to the drastic drop in number density (dashed curves in panels a4, b4), which suggests the presence of rarefied medium very far away from the central object . 

Shock signatures are present in the flow variables. 
They make the flow subsonic, represented by a lowering of velocity (solid curve in panel a1, dashed curve in b1). Temperature, on the other hand, jumps to a higher value due to compression (solid curve in panel a2, dashed curve in b2). This is directly correlated with the increase in number density (solid curve in panel a4, dashed curve in b4), which increases the thermal energy of the system. The compression ratio ($n_+/n_-$) is 1.773 for the accretion shock ($\lambda=1.5$) while it is 1.987 for wind shock ($\lambda=1.7$). This higher temperature is accompanied by a lowering of $\Gamma$ (solid curve in panel a3, dashed curve in b3). A higher angular momentum flow is accompanied by higher heating as well as cooling. This is because, a large amount of time is present for the matter to interact between themselves (because of higher $v_\phi$ and hence lower radial $v$). Hence, the flow temperature of $\lambda=1.7$ (panel b2) at any given $r$ is higher, than in case of $\lambda=1.5$ (panel a2). 

\section{Conclusions \& Future Work}
In the current work, we have proposed a novel methodology to obtain critical/sonic points, given any set of flow parameters. IRM-SP and IRM-SHOCK techniques have proved to successful in reproducing all class of global transonic solutions. These methods conserve all the natural properties of a flow, and use least possible assumptions to obtain a solution. It has been able to provide a unification between  accretion solutions and winds. IRM-SP is sufficient to obtain solutions covering almost whole of the parameter space, since all types of flows are connected to the central object. However, only in the case of wind shocks, IRM-SHOCK methodology is employed, since the non-global TS passing through $\rco$ does not connect to the surface. An explicit adaptive-RK method has been used for integration of the EoM in the IRM-SP and IRM-SHOCK technique, which increases efficiency of problem solving and optimizes speed.

One of the major limitations of this work is the use of a simplified $\alpha P$ prescription for viscosity. A more complicated and realistic prescription of viscous stress involves the gradient of angular velocity ($d\Omega/dr$) \citep{kc14,bl03}. This requires solving an additional differential equation for $\lambda$ (which is coupled with other differential equations), unlike in the current work, where an analytical expression of $\lambda\equiv\lambda(r,v,\Theta) $ was used. 
In the former prescription ($\trp\propto d\Omega/dr$), $d\lambda/dr$ equation cannot be integrated using the commonly-used integration techniques, like the explicit RK methods, as have been used in the current work. This was mentioned BL03. Thus, obtaining all TSs are impossible. Hence, we need to employ a different integration scheme and it will be discussed in an upcoming paper. However, the algorithm described in the current paper remains unaltered, except that the integration scheme only changes. This paper would hence, serve as a foundation. Also, we aim to connect the observational signatures, with the obtained solution spectra. This work is under preparation and the results would be reported in an upcoming work.

\section{Appendix: Stability of Cash-Karp method - an explicit adaptive Runge-Kutta method}
Cash-Karp (CK) method is used for solving ordinary differential equations (ODEs)\cite{ck}. It is a type of embedded Runge-Kutta method where six evaluations are needed to obtain a $5^{\rm th}$-order accurate solution.

\begin{itemize}
\item 
\textit{Adaptive step size}: In the family of ODE solvers, RK methods are known to be more efficient as well as accurate than the Euler method. However, they are difficult to implement, especially in case of coupled ODEs, where at each evaluation 3 EoM needs to be solved. Thus, RK4 method has been widely used to obtain accretion/wind solutions, because of its simplicity and less function evaluations, than its higher-order counterparts. Unfortunately, RK4 suffer from a fixed step-size issue. The solutions we deal with, have sharp gradients in velocity and temperature in certain regions, while they are smooth at other regions. Thus, a fixed step size is untenable. Also, since we are dealing with large radial distances ($1.001<r<10^4$), an unnaturally small step-size would require heavy computational time. Thus, for efficiency as well as accuracy we take recourse to the Cash-Karp method. This method computes the error by finding the difference between its $4^{\rm th}$ and  $5^{\rm th}$ order solution, which is  then required to determine the step-size. A big error will reduce the step-size and vice-versa\cite{press}. A user-defined tolerance level is set up, to limit the magnitude of error accepted by the system.

\item Stability analysis: Here, we investigate the stability  of CK method assuming a linear ODE, $y'= \lambda y$. The solution of this equation is simple, $y=\exp (\lambda x)$. Suppose, $z=h\lambda$ ($h$ being the step-size), then the solution converges for $z<0$. A single-step method for solving the above ODE is $y_{n+1}=R(z) y_n$, where $R(z)$ is the amplification factor to go to the next step, or it can also be called as the stability function for the particular single-step method. Since, the approximate solution is multiplied by this factor at each step, for the solution to converge or remain bounded, we must have $|R(z)\le  1|$. The region where $z$ satisfies the following condition $S := \lbrace z  \in {\mathbb C};|R(z)| \le 1\rbrace$  is called the region of stability. 
For example, in case of Euler's method, $y_{n+1}=y_n+hy'(x_n,y_n)=(1-h\lambda)y_n$.  Here, the method is stable when, $|R(z)|=|1-h\lambda|\le 1$. Similarly, we derive the stability regions of classical RK4 and the Cash-Karp methods and plot them in Fig.~\ref{fig:st}. The region  inside the contours satisfy $|R(z)| \le 1$, which suggests that the corresponding method is stable only for those values of $h\lambda$. 

\begin{figure}[h]
\centering
\includegraphics[scale=0.6]{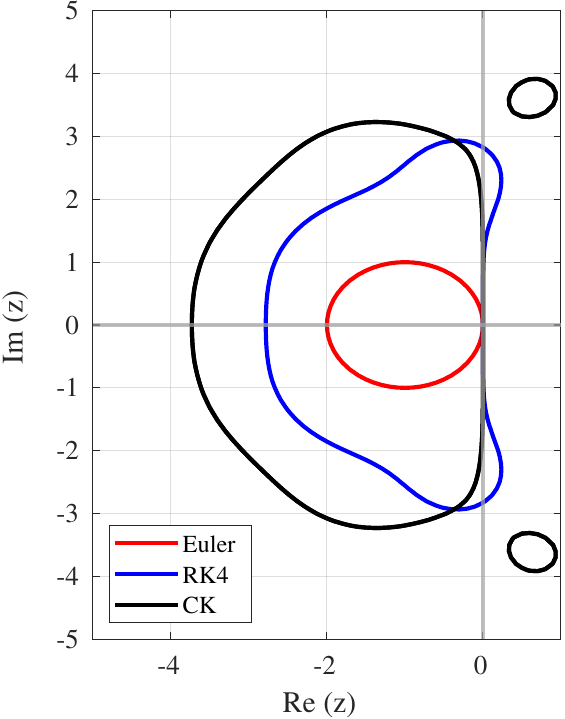}
\caption{Stability regions for Euler (red), classical RK4 (blue) and Cash-Karp (black) methods.}
\label{fig:st}
\end{figure}

\end{itemize}





\begin{thebibliography}{24}
\expandafter\ifx\csname natexlab\endcsname\relax\def\natexlab#1{#1}\fi
\expandafter\ifx\csname bibnamefont\endcsname\relax
  \def\bibnamefont#1{#1}\fi
\expandafter\ifx\csname bibfnamefont\endcsname\relax
  \def\bibfnamefont#1{#1}\fi
\expandafter\ifx\csname citenamefont\endcsname\relax
  \def\citenamefont#1{#1}\fi
\expandafter\ifx\csname url\endcsname\relax
  \def\url#1{\texttt{#1}}\fi
\expandafter\ifx\csname urlprefix\endcsname\relax\def\urlprefix{URL }\fi
\providecommand{\bibinfo}[2]{#2}
\providecommand{\eprint}[2][]{\url{#2}}

  
\bibitem[Becker and Le(2003)]{bl03}Becker, P.A. and Le, T.: 2003, {\it The Astrophysical Journal} {\bf 588}, 408. doi:10.1086/368377.

\bibitem[Frank, King, and Raine (2002)]{king}Frank, J., King, A., and Raine, D.J.: 2002, {\it Accretion Power in Astrophysics, by Juhan Frank and Andrew King and Derek Raine, pp. 398. ISBN 0521620538. Cambridge, UK: Cambridge University Press, February 2002.}, 398.
\bibitem[Chakrabarti(1996)]{c96}Chakrabarti, S.K.: 1996, {\it Physics Reports} {\bf 266}, 229. doi:10.1016/0370-1573(95)00057-7.

\bibitem[Hoyle and Lyttleton(1939)]{hl39}Hoyle, F. and Lyttleton, R.A.: 1939, {\it Proceedings of the Cambridge Philosophical Society} {\bf 35}, 405. doi:10.1017/S0305004100021150.
\bibitem[Bondi and Hoyle(1944)]{bh44}Bondi, H. and Hoyle, F.: 1944, {\it Monthly Notices of the Royal Astronomical Society} {\bf 104}, 273. doi:10.1093/mnras/104.5.273.

\bibitem[Bondi(1952)]{b52}Bondi, H.: 1952, {\it Monthly Notices of the Royal Astronomical Society} {\bf 112}, 195. doi:10.1093/mnras/112.2.195.
\bibitem[Shakura and Sunyaev(1973)]{ss73}Shakura, N.I. and Sunyaev, R.A.: 1973, {\it Astronomy and Astrophysics} {\bf 24}, 337.
\bibitem[Shapiro, Lightman, and Eardley(1976)]{sle76}Shapiro, S.L., Lightman, A.P., and Eardley, D.M.: 1976, {\it The Astrophysical Journal} {\bf 204}, 187. doi:10.1086/154162.

\bibitem[Ichimaru(1977)]{ichi77}Ichimaru, S.: 1977, {\it The Astrophysical Journal} {\bf 214}, 840. doi:10.1086/155314.

\bibitem[Abramowicz \emph{et al.}(1988)]{abra88}Abramowicz, M.A., Czerny, B., Lasota, J.P., and Szuszkiewicz, E.: 1988, {\it The Astrophysical Journal} {\bf 332}, 646. doi:10.1086/166683.
\bibitem[Abramowicz \emph{et al.}(1996)]{abra96}Abramowicz, M.A., Chen, X.-M., Granath, M., and Lasota, J.-P.: 1996, {\it The Astrophysical Journal} {\bf 471}, 762. doi:10.1086/178004.
\bibitem[Chakrabarti(1996)]{ch96}Chakrabarti, S.K.: 1996, {\it The Astrophysical Journal} {\bf 464}, 664. doi:10.1086/177354.

\bibitem[Chen, Abramowicz, and Lasota(1997)]{chen97}Chen, X., Abramowicz, M.A., and Lasota, J.-P.: 1997, {\it The Astrophysical Journal} {\bf 476}, 61. doi:10.1086/303592.
\bibitem[Narayan and Yi(1994)]{ny94}Narayan, R. and Yi, I.: 1994, {\it The Astrophysical Journal} {\bf 428}, L13. doi:10.1086/187381.


\bibitem[Event Horizon Telescope Collaboration \emph{et al.}(2019)]{eht1}Event Horizon Telescope Collaboration, Akiyama, K., Alberdi, A., Alef, W., Asada, K., Azulay, R., and, ...: 2019, {\it The Astrophysical Journal} {\bf 875}, L1. doi:10.3847/2041-8213/ab0ec7.
\bibitem[Event Horizon Telescope Collaboration \emph{et al.}(2019)]{eht2}Event Horizon Telescope Collaboration, Akiyama, K., Alberdi, A., Alef, W., Asada, K., Azulay, R., and, ...: 2019, {\it The Astrophysical Journal} {\bf 875}, L6. doi:10.3847/2041-8213/ab1141.
\bibitem[Event Horizon Telescope Collaboration \emph{et al.}(2022)]{eht3}Event Horizon Telescope Collaboration, Akiyama, K., Alberdi, A., Alef, W., Algaba, J.C., Anantua, R., and, ...: 2022, {\it The Astrophysical Journal} {\bf 930}, L12. doi:10.3847/2041-8213/ac6674.

\bibitem[Event Horizon Telescope Collaboration \emph{et al.}(2022)]{eht4}Event Horizon Telescope Collaboration, Akiyama, K., Alberdi, A., Alef, W., Algaba, J.C., Anantua, R., and, ...: 2022, {\it The Astrophysical Journal} {\bf 930}, L17. doi:10.3847/2041-8213/ac6756.


\bibitem[Manmoto, Mineshige, and Kusunose(1997)]{man97}Manmoto, T., Mineshige, S., and Kusunose, M.: 1997, {\it The Astrophysical Journal} {\bf 489}, 791. doi:10.1086/304817.
\bibitem[Mahadevan(1998)]{maha98}Mahadevan, R.: 1998, {\it Nature} {\bf 394}, 651. doi:10.1038/29241.

\bibitem[Nemmen \emph{et al.}(2006)]{nem06}Nemmen, R.S., Storchi-Bergmann, T., Yuan, F., Eracleous, M., Terashima, Y., and Wilson, A.S.: 2006, {\it The Astrophysical Journal} {\bf 643}, 652. doi:10.1086/500571.
\bibitem[Mo{\'s}cibrodzka \emph{et al.}(2009)]{monika09}Mo{\'s}cibrodzka, M., Gammie, C.F., Dolence, J.C., Shiokawa, H., and Leung, P.K.: 2009, {\it The Astrophysical Journal} {\bf 706}, 497. doi:10.1088/0004-637X/706/1/497.
\bibitem[Yuan and Narayan(2014)]{yn14}Yuan, F. and Narayan, R.: 2014, {\it Annual Review of Astronomy and Astrophysics} {\bf 52}, 529. doi:10.1146/annurev-astro-082812-141003.

\bibitem[Bisnovatyi-Kogan and Lovelace(2001)]{bkl01}Bisnovatyi-Kogan, G.S. and Lovelace, R.V.E.: 2001, {\it New Astronomy Reviews} {\bf 45}, 663. doi:10.1016/S1387-6473(01)00146-4.

\bibitem[Narayan and Yi(1995)]{ny95}Narayan, R. and Yi, I.: 1995, {\it The Astrophysical Journal} {\bf 452}, 710. doi:10.1086/176343.
\bibitem[Honma \emph{et al.}(1991)]{honma91}Honma, F., Matsumoto, R., Kato, S., and Abramowicz, M.A.: 1991, {\it Publications of the Astronomical Society of Japan} {\bf 43}, 261.
\bibitem[Chen and Taam(1993)]{chentaam}Chen, X. and Taam, R.E.: 1993, {\it The Astrophysical Journal} {\bf 412}, 254. doi:10.1086/172916.


\bibitem[Liang and Thompson(1980)]{lt80}Liang, E.P.T. and Thompson, K.A.: 1980, {\it The Astrophysical Journal} {\bf 240}, 271. doi:10.1086/158231.
\bibitem[Fukue(1987)]{f87}Fukue, J.: 1987, {\it Publications of the Astronomical Society of Japan} {\bf 39}, 309.
\bibitem[Kafatos and Yang(1994)]{kaf94}Kafatos, M. and Yang, R.X.: 1994, {\it Monthly Notices of the Royal Astronomical Society} {\bf 268}, 925. doi:10.1093/mnras/268.4.925.
\bibitem[Lee(1999)]{lee99}Lee, U.: 1999, {\it The Astrophysical Journal} {\bf 511}, 359. doi:10.1086/306659.
\bibitem[Kumar and Chattopadhyay(2014)]{kc14}Kumar, R. and Chattopadhyay, I.: 2014, {\it Monthly Notices of the Royal Astronomical Society} {\bf 443}, 3444. doi:10.1093/mnras/stu1389.


\bibitem[Sarkar, Chattopadhyay, and Laurent(2020)]{scp20}Sarkar, S., Chattopadhyay, I., and Laurent, P.: 2020, {\it Astronomy and Astrophysics} {\bf 642}, A209. doi:10.1051/0004-6361/202037520.

\bibitem[Sarkar \emph{et al.}(2023)]{skcp23}Sarkar, S., Singh, K., Chattopadhyay, I., and Laurent, P.: 2023, {\it Monthly Notices of the Royal Astronomical Society} {\bf 522}, 3735. doi:10.1093/mnras/stad1064.

\bibitem[Paczy{\'n}sky and Wiita(1980)]{pw80}Paczy{\'n}sky, B. and Wiita, P.J.: 1980, {\it Astronomy and Astrophysics} {\bf 88}, 23.

\bibitem[Chattopadhyay and Ryu(2009)]{cr09}Chattopadhyay, I. and Ryu, D.: 2009, {\it The Astrophysical Journal} {\bf 694}, 492. doi:10.1088/0004-637X/694/1/492.

\bibitem[Sarkar and Chattopadhyay(2019)]{sc19ijmpd}Sarkar, S. and Chattopadhyay, I.: 2019, {\it International Journal of Modern Physics D} {\bf 28}, 1950037. doi:10.1142/S0218271819500378.

\bibitem[Sarkar and Chattopadhyay(2022)]{sc22jaa}Sarkar, S. and Chattopadhyay, I.: 2022, {\it Journal of Astrophysics and Astronomy} {\bf 43}, 34. doi:10.1007/s12036-022-09820-z.

\bibitem[Sarkar and Chattopadhyay(2019)]{sc19jp}Sarkar, S. and Chattopadhyay, I.: 2019, {\it Journal of Physics Conference Series} {\bf 1336}, 012019. doi:10.1088/1742-6596/1336/1/012019.


\bibitem[Sarkar and Chattopadhyay(2020)]{sc20jp}Sarkar, S. and Chattopadhyay, I.: 2020, {\it Journal of Physics Conference Series} {\bf 1640}, 012022. doi:10.1088/1742-6596/1640/1/012022.
\bibitem[Becker and Subramanian(2005)]{bs05}Becker, P.A. and Subramanian, P.: 2005, {\it The Astrophysical Journal} {\bf 622}, 520. doi:10.1086/427769.

\bibitem[S{{a}}dowski(2009)]{sado09}S{{a}}dowski, A.: 2009, {\it The Astrophysical Journal Supplement Series} {\bf 183}, 171. doi:10.1088/0067-0049/183/2/171.


\bibitem[Matsumoto \emph{et al.}(1984)]{mat84}Matsumoto, R., Kato, S., Fukue, J., and Okazaki, A.T.: 1984, {\it Publications of the Astronomical Society of Japan} {\bf 36}, 71.


\bibitem[Nakamura \emph{et al.}(1996)]{naka96}Nakamura, K.E., Matsumoto, R., Kusunose, M., and Kato, S.: 1996, {\it Publications of the Astronomical Society of Japan} {\bf 48}, 761. doi:10.1093/pasj/48.5.761
\bibitem[Datta, Mondal, and Mukhopadhyay(2022)]{sudeb22}Datta, S.R., Mondal, T., and Mukhopadhyay, B.: 2022, {\it Monthly Notices of the Royal Astronomical Society} {\bf 513}, 204. doi:10.1093/mnras/stac835.
\bibitem[Maity \emph{et al.}(2022)]{susovan22}Maity, S., Shaikh, M.A., Tarafdar, P., and Das, T.K.: 2022, {\it Physical Review D} {\bf 106}, 044062. doi:10.1103/PhysRevD.106.044062.
\bibitem[Landau and Lifshitz(1959)]{ll59}Landau, L.D. and Lifshitz, E.M.: 1959, {\it Course of theoretical physics, Oxford: Pergamon Press, 1959}.


\bibitem[Popham and Narayan(1992)]{gn92}Popham, R. and Narayan, R.: 1992, {\it The Astrophysical Journal} {\bf 394}, 255. doi:10.1086/171577.
\bibitem[Mitra \emph{et al.}(2023)]{mitra23}Mitra, S., Ghoreyshi, S.M., Mosallanezhad, A., Abbassi, S., and Das, S.: 2023, {\it Monthly Notices of the Royal Astronomical Society} {\bf 523}, 4431. doi:10.1093/mnras/stad1682.
\bibitem[Bu, Qiao, and Yang(2019)]{bu19}Bu, D.-F., Qiao, E., and Yang, X.-H.: 2019, {\it The Astrophysical Journal} {\bf 875}, 147. doi:10.3847/1538-4357/ab12ea.
\bibitem[J. R. Cash and Alan H. Karp(1990)]{ck}J. R. Cash and Alan H. Karp.: 1990, ACM Trans. Math. Softw. 16, 3 (Sept. 1990), 201-222. https://doi.org/10.1145/79505.79507
\bibitem[Press, Flannery, and Teukolsky(1986)]{press}Press, W.H., Flannery, B.P., and Teukolsky, S.A.: 1986, {\it Cambridge: University Press, 1986}.


\end{thebibliography}
\end{document}